\documentclass[reprint,pra,showpacs,preprintnumbers,amsmath,amssymb,aps,superscriptaddress]{revtex4-1}
\usepackage{graphicx}

\usepackage{subcaption}
\usepackage{dcolumn}
\usepackage{amsmath}
\usepackage{times}
\usepackage[braket, qm]{qcircuit}
\usepackage{physics}
\usepackage{mathtools}
\usepackage{wasysym}
\usepackage{caption}
\usepackage{booktabs}
\begin{document}

\title{Comparing Zeeman qubits to hyperfine qubits in the context of the surface code: $^{174}$Yb$^{+}$ and $^{171}$Yb$^{+}$}

\author{Natalie C. Brown}
\affiliation{School of Physics,
Georgia Institute of Technology, Atlanta, GA, USA}
\author{Kenneth R. Brown}
\affiliation{School of Physics,
Georgia Institute of Technology, Atlanta, GA, USA}
\affiliation{Schools of Chemistry and Biochemistry and Computational Science and Engineering, Georgia Institute of Technology, Atlanta, GA, USA}
\affiliation{Departments of Electrical and Computer Engineering, Chemistry, and Physics, Duke University, Durham, NC, USA }

\date{\today}

\begin{abstract}
Many systems used for quantum computing possess additional states beyond those defining the qubit. Leakage out of the qubit subspace must be considered when designing quantum error correction codes.  Here we consider trapped ion qubits manipulated by Raman transitions.  Zeeman qubits do not suffer from leakage errors but are sensitive to magnetic fields to first-order.  Hyperfine qubits can be encoded in clock states that are insensitive to magnetic fields to first-order, but spontaneous scattering during the Raman transition can lead to leakage. Here we compare a Zeeman qubit ($^{174}$Yb$^+$) to a hyperfine qubit ($^{171}$Yb$^+$) in the context of  the surface code.  We find that the number of physical qubits required to reach a specific logical qubit error can be reduced by using $^{174}$Yb$^+$ if the magnetic field can be stabilized  with fluctuations smaller than $10$ $\mu$G.
\end{abstract}
\maketitle
\section{Introduction}
\label{intro}
The ideal qubit consists of a pair of orthogonal quantum states. However most systems used for quantum computing (QC) are multilevel systems and these additional levels allow for leakage out of the qubit subspace. Leakage errors result in the quantum system leaving the computational space and are suffered by trapped ions \cite{duan2001geometric, haffner2008quantum, cirac1995quantum, plenio1997decoherence}, quantum dots \cite{byrd2005universal, fong2011universal, mehl2015fault}, superconducting qubits \cite{zhou2005rapid, motzoi2009simple, ferron2010intrinsic, herrera2013tradeoff, ghosh2013understanding} and anyons \cite{xu2008constructing, ainsworth2011topological}.

Because leakage faults occur outside the computational space, traditional methods for correcting Pauli type errors are ineffective on them. Instead, the issue of leakage requires a separate set of techniques for reducing the faults. At the physical level, leakage errors can be mitigated through the use of different pulse techniques \cite{motzoi2009simple, ferron2010intrinsic, mcconkey2017mitigating, chen2016measuring}. Leakage errors can also be detected and converted to Pauli or erasure errors by constructing suitable leakage reducing units (LRUs) \cite{byrd2005universal, byrd2004overview, wu2002efficient, suchara,fowler, fowler2012surface, ghosh2013understanding, ghosh2015leakage}.  It is also possible to construct a system that does not suffer from leakage \cite{lucas2004isotope}. Thus when designing the architecture of a quantum computer is it worthwhile to examine the resources needed to deal with leakage.

Ion trapped computers are a leading candidate for QC \cite{haffner2008quantum}. Quantum information is encoded in the internal states of the ion, often a pair of levels in the $S_{1/2}$ ground state. The two states are connected by a magnetic dipole transition with  a small frequency difference, typically a radio-frequency for Zeeman qubits and a microwave frequency for hyperfine qubits, resulting in a practically infinite lifetime of the excited level due to spontaneous decay \cite{wineland, toolbox, ozeri}. In ions with $I=0$, the only $S_{1/2}$ levels available are that of the two Zeeman states. Zeeman qubits do not suffer from leakage in the ground manifold states, but have a first order dependence on magnetic fields \cite{lucas2004isotope, keselman2011high, poschinger2009coherent, ratschbacher2013decoherence}.  In ions with $I \neq 0$, the qubit can be encoded into any pair of hyperfine states. However, the existence of other hyperfine states means there is a potential for leakage. Hyperfine qubits based on clock-states, have a second order dependance on magnetic fields but spontaneous scattering during stimulated Raman processes can lead to leakage errors \cite{brown2011single, ballance2016high, blinov2004quantum, olmschenk2007manipulation}. 

Two-photon Raman transitions are often used to manipulate qubits in ion traps \cite{wineland, toolbox, ozeri, haffner2008quantum, lucas2004isotope, ratschbacher2013decoherence, poschinger2009coherent, brown2011single}. Quantum gates rely on coupling to excited states through electric dipole transitions. Since laser light is used to drive these transitions, spontaneous scattering of photons is inevitable. While detuning the laser frequency away from allowed optical transitions can suppress this scattering, it is impossible to completely eliminate. Both Raman and Rayleigh scattering can lead to decoherence but each manifest differently depending on qubit choice \cite{ozeri, uys, ozeri2005hyperfine}.  We note that scattering errors can be avoided by using only microwave gates \cite{HartyPRL2016, WeidtPRL2016, KhromovaPRL2012, OspelkausPRL2008}, but leakage due to background gas collisions or imperfections in operations could still occur.

This work seeks to quantify these errors in the context of quantum error correction (QEC). First we describe the characteristics associated with each type of qubit as well as their magnetic field dependence. Next we discuss the calculation of the different errors associated with spontaneous scattering from driven Raman transitions. Finally we compare the ions in the context of the surface code. Our results show leakage is more prominent than expected, and given a stable enough magnetic field, Zeeman qubits require a smaller distance surface code to produce the same logical error rate as a logical qubit composed from a physical hyperfine qubit. 

\section{Yb$^+$ Model and Associated Errors}
\begin{figure}[h]
\includegraphics[width=7cm, height=5cm]{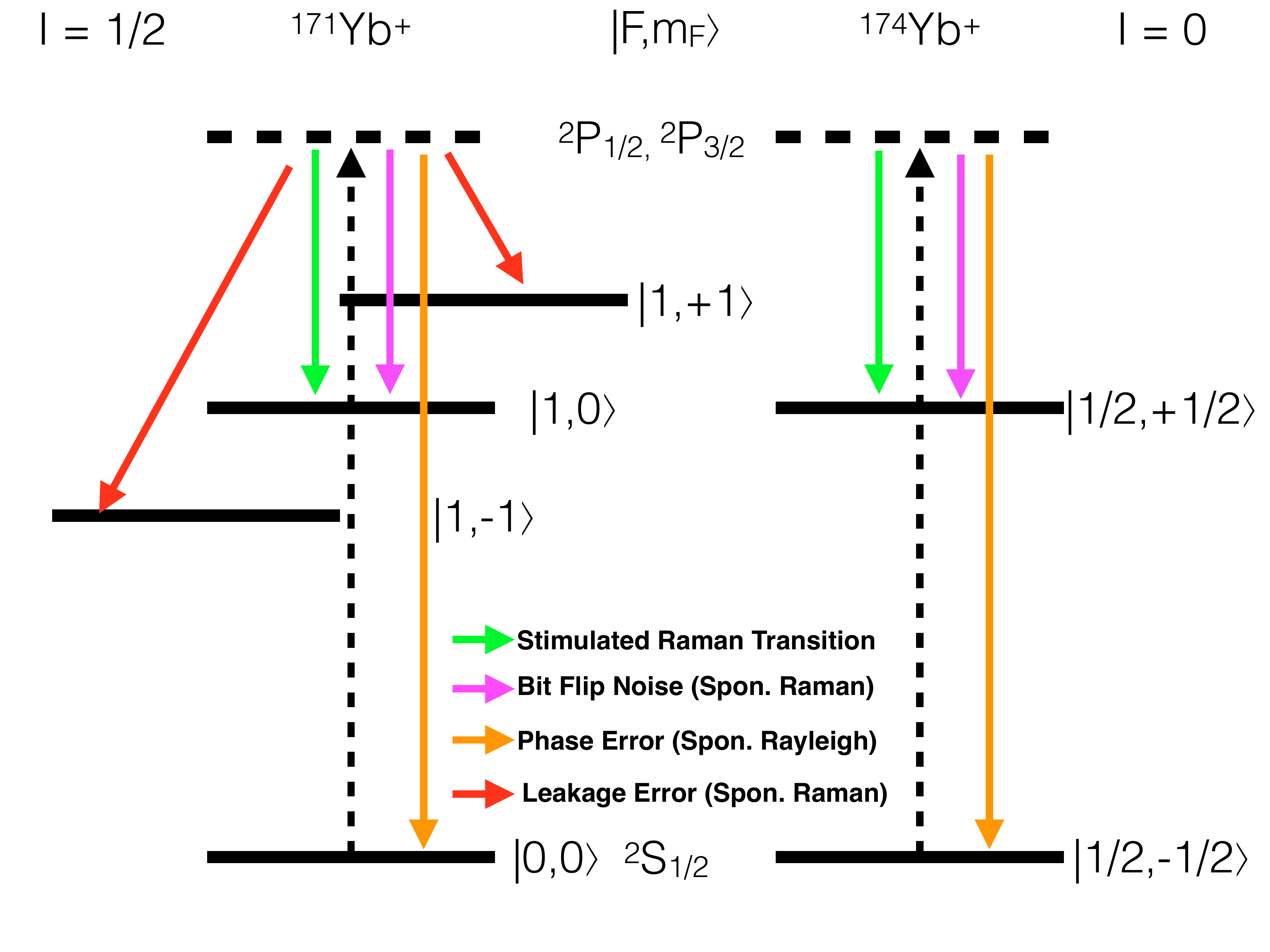}
\caption{Atomic structure of Yb$^+$ isotopes and errors associated with different scattering events from the $^{2}$P states assuming the ion starts in the lower qubit state. Spontaneous Raman scattering can cause bit flip noise or leakage errors. Spontaneous Rayleigh scattering can lead to dephasing errors.}
\label{YbModel}
\end{figure}

Yb$^+$ has many naturally occurring isotopes but we examine, $^{174}$Yb$^+$ ($I=0$) and $^{171}$Yb$^+$ ($I=1/2$), whose nuclear spin yield a Zeeman and hyperfine qubit, respectively.  This makes Yb$^+$ the perfect candidate to study the associated error rates between these two types of qubits. The atomic structures and associated possible errors resulting from spontaneous scattering for both isotopes are illustrated in Fig. \ref{YbModel}. While there are other sources of noise that could be considered, we choose to focus on two types of noise that are the most relevant to the comparison of the two types of qubits: magnetic field fluctuations that lead to dephasing in  Zeeman qubits and spontaneous scattering that lead to leakage errors in hyperfine qubits. 

\subsection{Unstable Magnetic Field}
For the Zeeman qubit, $^{174}$Yb$^+$, the qubit is encoded into the electron spin states $\ket{S=1/2,m_s=-1/2}$ and $\ket{S=1/2, m_s=1/2}$. While there is no possibility for leakage (in this discussion we assume higher-level leakage states in the D and F manifolds are quickly repumped to the ground state), because the qubit itself is encoded in Zeeman energy splitting, it will be highly susceptible to magnetic field fluctuations. The applied magnetic field required for the ion trap causes the well known Zeeman energy splitting and the first order effects grow linearly with the magnetic field. Any deviations in the magnetic field yield a first order frequency shift given by
 \begin{equation}
\Delta \nu = \frac{g_s \mu_B }{\hbar} \Delta B
\end{equation} 
where $g_s$ is the Land\'{e} \textit{g}-factor, $\mu_B$ is the Bohr magneton, $\hbar$ is Planck's constant and $\Delta B$ is the difference from between the actual magnetic field and the ideal magnetic field \cite{PhysRev.49.324}.  Such magnetic field noise can cause dephasing and is the main disadvantage of using a Zeeman qubit. 

For $^{171}$Yb$^+$, the qubit is encoded into the clock states $\ket{F=1,m_F=0}$ and $\ket{F=0,m_F=0}$. These states are magnetic field insensitive transitions that do not suffer from first order effects. The second order magnetic field dependence can be derived from the Briet-Rabi formula with the frequency shift due to uncertainties in the magnetic field given by 
\begin{equation}
\Delta \nu = \frac{(g_J - g_I)^2 \mu_B^2 }{2 \hbar^2 \omega} (2B_0\Delta B +(\Delta B)^2)
\end{equation}
where $g_J$ and $g_I$ are the Land\'{e} \textit{g}-factors for the electron and the nucleus, $\omega$ is the angular frequency of the hyperfine splitting, $B_0$ is the ideal magnetic field strength, and $\Delta B$ is the deviation from the ideal magnetic field~\cite{PhysRev.49.324, zeeman}. Because the second order effect is so small, clock states are negligibly affected by magnetic field noise, a clear advantage when using hyperfine qubits. At typical values of applied magnetic fields for hyperfine qubits, the effective frequency fluctation is 10$^{-3}$ to 10$^{-4}$ smaller than for the Zeeman qubit.  However, the existence of the other hyperfine states $\ket{1,+1}$ and $\ket{1, -1}$ in $^{171}$Yb$^{+}$ can lead to leakage events. 

Using equations $(1)$ and $(2)$, we assumed a Gaussian distribution and calculated the probability of error based on gate time and magnetic field stability. For low errors, the error from the first order Zeeman effect grows quadratically with increasing field fluctuations. For fields with high fluctuations, this probability is well above the threshold error value of the surface code of ~1$\%$ \cite{dennis2002topological, raussendorf2007fault}.  
The probability of error resulting from the second order effects grows quartically with field fluctuations in the limit of zero average magnetic field. Even at fields with low stability, this error remains below threshold. Table \ref{table1} lists these probabilities with varying magnetic field stabilities for both single and two-qubit $\hat{I}$ gates. The more stable the field, the less error. The errors vary drastically for Zeeman qubits and are almost negligible for hyperfine qubits.

\begin{table}
\resizebox{\columnwidth}{!}{%
  \begin{tabular}{|c|c|c|c|c|}
    \hline
     &
    \multicolumn{2}{|c|}{Single Qubit Gate}&
    \multicolumn{2}{|c|}{Two-Qubit Gate}\\
  
    &
    \multicolumn{2}{|c|}{$\tau_{gate}=1$ $\mu$s}&
    \multicolumn{2}{|c|}{$\tau_{gate}=200$ $\mu$s}\\
     \hline
    Probability & $^{171}$Yb$^{+}$ & $^{174}$Yb$^{+}$ &  $^{171}$Yb$^{+}$ &$ ^{174}$Yb$^{+}$  \\ \hline
      ${P_{\sigma=10^{-2}}}$ &$1.90\times10^{-14}$&$1.93\times 10^{-3}$ & $7.62\times10^{-10}$ & $0.50$ \\
    \hline
      ${P_{\sigma=10^{-3}}}$ &$1.90\times10^{-18}$&$1.93\times 10^{-5}$ & $7.62\times10^{-14}$ & $0.39$ \\
    \hline
      ${P_{\sigma=10^{-4}}}$ &$1.90\times10^{-22}$&$1.93 \times 10^{-7}$ & $7.62\times10^{-18}$ & $7.69 \times 10^{-3}$ \\
    \hline
      ${P_{\sigma=10^{-5}}}$ &$1.90\times10^{-26}$&$1.93\times10^{-9}$ & $7.62\times10^{-22}$ & $7.75\times 10^{-5}$ \\
    \hline
      ${P_{\sigma=10^{-6}}}$ &$1.90\times10^{-30}$&$1.93\times 10^{-11}$ & $7.62\times10^{-26}$ & $7.75\times 10^{-7}$ \\
    \hline
  \end{tabular}
  }\caption{A list of error probabilities caused by the first order Zeeman effect ($^{174}$Yb$^+$) and the second order Zeeman effect ($^{171}$Yb$^+$). The gate times for one and two-qubits gates were 1 $\mu$s and 200 $\mu$s, respectively. $\sigma$ is the standard deviation of the magnetic field strength  in G. The table shows $^{171}$Yb$^+$ error for zero average magnetic field.  For typical magnetic fields yielding 1 MHz Zeeman splittings, the error for  $^{171}$Yb$^+$ for a given $\sigma$ is comparable to the error for  $^{174}$Yb$^+$ with $\sigma^\prime  = 10^{-4}  \sigma $.  }
\label{table1}
\end{table}

\subsection{Spontaneous Scattering}
Additional errors arise from the scattering of photons during gates. Two-photon Raman coupling is among the most popular choices for gate implementation \cite{haffner2008quantum, lucas2004isotope, poschinger2009coherent, brown2011single, ballance2016high, blinov2004quantum, olmschenk2007manipulation, wineland, ozeri, uys}. Lasers detuned off-resonance drive qubit transitions through interactions with excited states. This use of stimulated transition to perform a qubit rotation lends itself to spontaneous emission errors. Raman scattering is usually thought of as the biggest contributor to these errors as all qubit types suffer from it \cite{ozeri}. Spontaneous Raman scattering can lead to leakage errors, or change the qubit in the computational basis ($\hat{X}$/$\hat{Y}$ error). Unlike leakage errors, Pauli type errors can be corrected using standard quantum error correction codes (QECC). Rayleigh scattering is typically less of a contributor to errors as it does not necessarily cause decoherence in all qubit types and in certain cases can be ignored \cite{ozeri,  uys, ozeri2005hyperfine}. Rayleigh scattering leads to dephasing errors ($\hat{Z}$), similar to the magnetic field fluctuations. This decoherence rate is dependent on the scattering amplitudes of the qubit levels and thus varies from isotope to isotope. 

To calculate the different error rates for the two ions, we followed the procedure outlined in \cite{uys}. The rate at which the ion in state $\ket{i}$ scatters a photon and ends in state $\ket{j}$ is given by the Kramers-Heisenberg formula
\begin{equation}
\Gamma_{ij} =(\frac{\mu E_0 }{2\hbar})^2 \gamma \sum_{\lambda}(\sum_{J}A^{i \rightarrow j}_{J,\lambda})^2
\end{equation}
where $\mu$ is the largest element of the dipole matrix, $E_0$ is the magnitude of a nonresonant light field of the lasers, $\gamma$ is the spontaneous decay rate of the excited states and $A^{i \rightarrow j}_{J,\lambda}$ are the scattering amplitudes \cite{uys, ozeri}.

The total Raman and effective Rayleigh scattering rates are given by
\begin{equation}
 \Gamma_{Ram} = \Gamma_{ij}+\Gamma_{ji}
 \end{equation}
  \begin{equation}
 \Gamma_{el} = (\frac{\mu E_0}{2\hbar})^2 \gamma \sum_{\lambda}(\sum_J A^{j \rightarrow j}_{J,\lambda} - \sum_{J^\prime}A_{J^\prime,\lambda}^{i \rightarrow i})^2
 \end{equation} 
respectively, where $i$ and $j$ represent the two qubit levels.  When Rayleigh scattering rates from the two ion qubit states are different, the scattered photons will measure the qubit state causing decoherence \cite{uys}. Thus the effective Rayleigh scattering that will cause dephasing is given by this difference. 
We calculated fidelity for both single ($\tau = 1$ $\mu$s) and two-qubit ($\tau = 200$ $\mu$s) gates of a $\pi$ rotation about the x-axis on the Bloch sphere. These gates were assumed to be driven by co-propogating linearly polarized Raman beams, blue detuned from the $P_{1/2}$ level with laser frequency of $355$ nm and a beam waist $w_0 = 20$ $\mu$m. The choice of these parameters was motivated by desired gate times, the minimization of spontaneous scattering and by recent experiments performed using a $355$ nm laser \cite{linke2017fault, leung2018robust, debnath2016demonstration, fallek2016transport}.  

Table \ref{table2} shows the different scattering errors for both the $^{174}$Yb$^+$ Zeeman and  $^{171}$Yb$^+$ hyperfine qubit. When the Rayleigh scattering amplitudes of the two qubit levels are approximately equal, their contributions can add up destructively. The decoherence rate due to Rayleigh scattering will be small and decoherence will be dominated by Raman scattering \cite{uys}. This is precisely what we see for $^{171}$Yb$^+$. However, even when amplitudes are approximately equal, they can have opposite sign and their different contributions can add up constructively leading to large Rayleigh scattering decoherence \cite{uys}, as in the case for $^{174}$Yb$^+$.  For $^{174}$Yb$^+$, Rayleigh scattering was approximately equal to the Raman scattering. In this sense, $^{174}$Yb$^+$ can be modeled anisotropically, with double the amount of Pauli $\hat{Z}$ type errors for every single Pauli $\hat{X}$ or $\hat{Y}$ type error. For $^{171}$Yb$^+$, Raman scattering that resulted in leakage was equal to the scattering which caused Pauli type errors. 

When looking at overall error rates, it is clear that a single $^{171}$Yb$^+$ is prone to less physical error. However, this hides the fact that leakage errors can be damaging to QECC. A majority of the errors that occur via spontaneous scattering in  $^{171}$Yb$^+$ (leakage errors) requires extra overhead to correct relative to pure Pauli errors. To gain a better understanding of this, we must look at how each type of qubit performs with a QECC.

\begin{table}

\resizebox{\columnwidth}{!}{%

  \begin{tabular}{|c|c|c|c|c|}
    \hline
   
     &
    \multicolumn{2}{|c|}{Single Qubit Gate}&
    \multicolumn{2}{|c|}{Two-Qubit Gate}\\
    
    &
     \multicolumn{2}{|c|}{$\tau_{gate}=1$ $\mu$s}&
    \multicolumn{2}{|c|}{$\tau_{gate}=200$ $\mu$s}\\
     \hline
    Probability & $^{171}$Yb$^{+}$ & $^{174}$Yb$^{+}$ &  $^{171}$Yb$^{+}$ &$ ^{174}$Yb$^{+}$  \\ \hline
    $P_{Raman}$ &$2.42\times10^{-6}$ & $4.8\times10^{-6}$ & $6.37\times 10^{-5}$ & $12.6 \times 10^{-5}$ \\
    \hline
    $P_{Leakage}$ &$2.42\times10^{-6}$  & N/A &$6.37\times 10^{-5}$ & N/A \\
    \hline
    $P_{Rayleigh}$ &$1.60\times10^{-13}$ &$4.88\times10^{-6}$ & $4.21\times10^{-12}$ & $12.6 \times 10^{-5}$ \\
    \hline
    \end{tabular}
   }
   \caption{A list of error probabilities caused by spontaneous scattering from stimulated Raman transitions. The gate times for one and two-qubits gates were 1 $\mu$s and 200 $\mu$s. The gates were assumed to by driven by co-propogating linearly polarized Raman beams with $f = 355$ nm and a beam waist of $w_0 = 20$ $\mu$m. For $^{174}$Yb$^+$, Rayleigh scattering was just as dominant as Raman scattering. For $^{171}$Yb$^+$, Raman scattering which resulted in leakage was equal to that of bit flip noise. }
\label{table2}
\end{table}

\section{Surface Code Model and LRC}
\begin{figure}[h]
\includegraphics[trim=250 450 250 125, width=2cm, height=3.5cm]{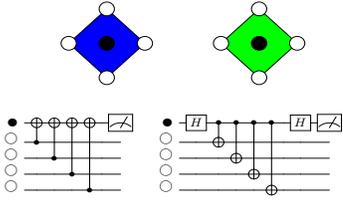}
\caption{Standard circuits to measure surface code check operators. The open white circles represent data qubits while the closed dark circles represent measure/ancilla qubits. The blue and green diamonds represent $\hat{Z}$ and $\hat{X}$ stabilizers respectively. }
\label{surface}
\end{figure}

The toric code was the first example of a topological code and is well studied \cite{kitaev2002classical, kitaev1997quantum, PhysRevA.90.032326}. The toric code is a two dimensional surface code with periodic boundary conditions and thus has a natural mapping onto the surface of a torus. Qubits are positioned in an array and either function as data qubits or ancilla/measurement qubits. Data qubits are used to encode the information while ancilla qubits are used to measure stabilizers, which in turn help infer where errors occurred. 
A six step cycle is implemented in order to perform one round of error correction. First, all ancilla qubits are initialized in their respective eigen basis (either $\ket{0}$ for $\hat{Z}$ or $\ket{+}$ for $\hat{X}$). Next, four CNOT gates are performed between each ancilla and data qubit. Finally, each ancilla is measured in it's respective basis. This is precisely the circuit outlined in Fig. \ref{surface}. The problem of inferring the most probable error given the observed syndrome is mapped to a minimum weight perfect matching problem that can be solved with Edmond's algorithm \cite{suchara}. Such error correcting schemes have been studied both with and without leakage \cite{ghosh2013understanding, ainsworth2011topological, fowler2012surface, fowler2012topological, suchara, fowler, ghosh2015leakage, PhysRevA.90.032326}.

\begin{figure}
\includegraphics[trim=110 425 20 115, clip, height=4cm, width=7cm]{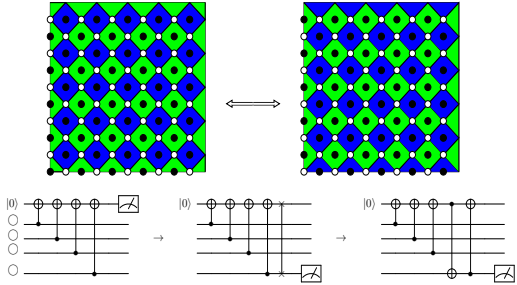}
\caption{The QUICK LRC required to perform error detection in the presence of leakage. After each cycle, the physical qubits get swaped. Data qubits become ancilla and ancilla qubits become data qubits. The information is transferred and leaked qubits get measured and reset every other cycle \cite{suchara}.}
\label{LRC}
\end{figure}

This six step error correction cycle is all that is needed to correct Pauli type errors. Handling leakage errors requires the use of LRU's. The idea of incorporating LRU's was first used to show an accuracy threshold exists even in the presence of leakage errors \cite{aliferis2005fault}. The most common type of LRU implements gate teleportation in some fashion \cite{aliferis2005fault, byrd2005universal, byrd2004overview, wu2002efficient}. The additional circuitry required to perform the teleportation is referred to as a leakage reducing circuit or LRC. Different strategies for implementing LRCs into surface codes have been studied \cite{aliferis2005fault, ghosh2013understanding, suchara, fowler, ghosh2015leakage}, in order to grasp the tradeoff between circuit complexity and effectiveness of leakage reduction. 

In our work, we chose to implement the Quick LRC \cite{suchara}, as depicted in Fig. \ref{LRC}. The Quick LRC adds a SWAP gate after the last CNOT of the standard circuit. At the end of each cycle, the physical qubits trade roles. Data qubits become ancilla qubits and ancilla qubits become data qubits. The cycle starts again reinitializing ancilla qubits. Leaked data qubits now get measured and reinitialized as ancilla qubits, and thus leaked qubits do not live for more than two cycles with this LRC implemented. Through the use of gate identities and gate cancellation, the implementation of this LRC requires only one additional CNOT. The Quick LRC is the simplest of all current LRCs and was shown to produced comparable results to that of more complicated LRCs \cite{suchara}. Other LRCs require more SWAP gates per cycle but did not show significant improvement compared to the QUICK LRC. In short, the Quick LRC effectively reduces leakage using the smallest overhead. 

\section{Results and Discussion}
Using the error probabilities calculated from the magnetic field fluctuations and the spontaneous scattering rates, we analyzed the performance of the Zeeman and hyperfine qubits on the toric code. The Zeeman qubit was demonstrated on the standard circuit (Fig. \ref{surface}) while the hyperfine qubit was demonstrated on the Quick LRC (Fig. \ref{LRC}).

In our model, after every gate magnetic field noise was introduce with probabilities corresponding to the magnetic field susceptibility of the qubits (Table \ref{table1}). Additionally, spontaneous scattering errors occurred after every gate with the ratios of the probability for a particular error corresponding to the calculated spontaneous scattering rate of the qubits (Table \ref{table2}), e.g. leakage was twice as probable as a Pauli $\hat{X}$/$\hat{Y}$, with the total probability of an scattering event equal to p. Spontaneous scattering also allows leaked qubits  to return to the qubit subspace.  The two qubits involved in a CNOT gate have independent probabilities to leak after the gate. Once the qubit leaked, it would remain leaked until a spontaneous scattering event returns it to the qubit subspace or the qubit is reset by the Quick LRC. While this means a leaked qubit was corrected at maximum every other error correction cycle, long lived leaked qubits had the potential of corrupting other qubits. 

When a CNOT is performed between a leaked qubit and a qubit in the computational basis, the latter suffers a random single-qubit Pauli error (including the trivial error $\hat{I}$), with equal probability. When a CNOT is performed between two qubits in the computational basis, the standard error propagation rules are applied. Magnetic field noise and spontaneous scattering errors are only applied after the gates to model environmental noise. Finally when a leakage qubit is measured, it yields a $\ket{+1}$ eigenvalue. This is physically motivated by the atomic structure of $^{171}$Yb$^+$ because any leaked state will be in the $F=1$ manifold and will be detected as such (see Fig. \ref{YbModel}). 
\begin{figure}[h]
\includegraphics[width=\columnwidth]{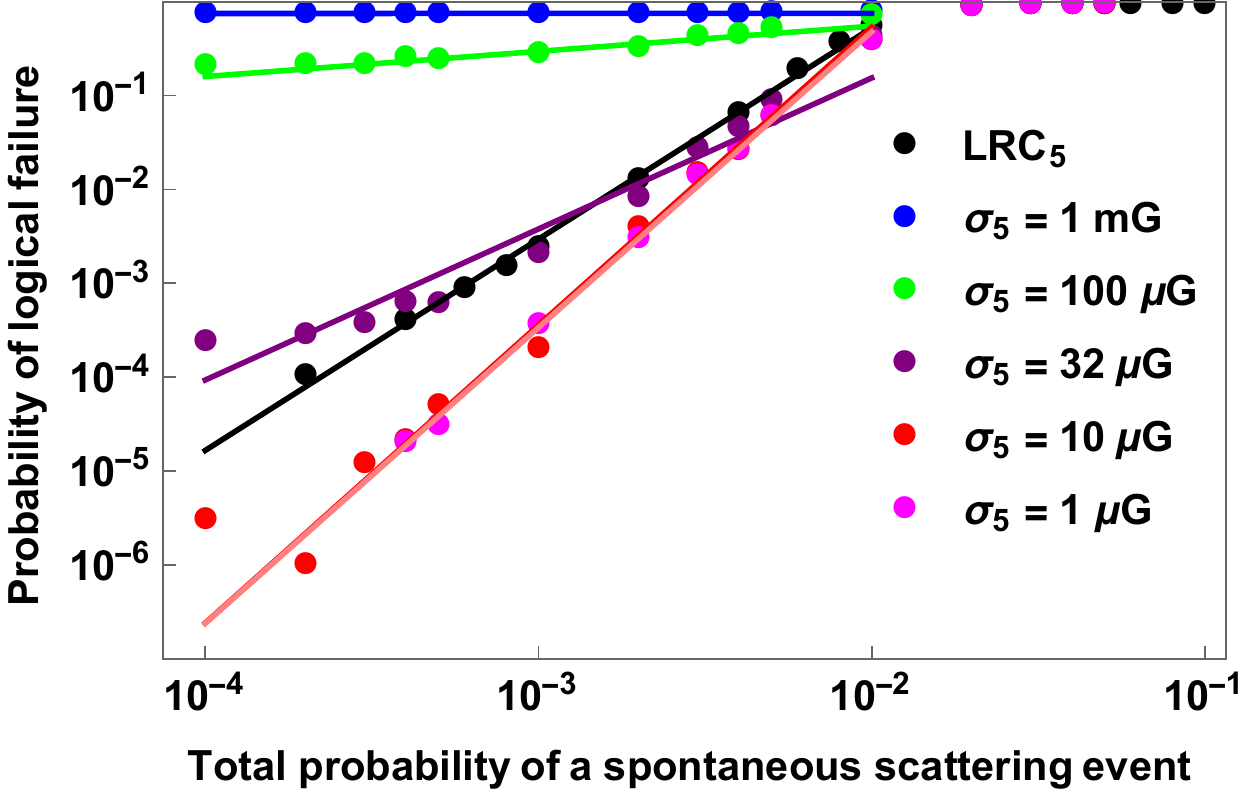}
\caption{Comparison of various magnetic field stabilities for a distance 5 code per 2 qubit gate. The hyperfine qubits (black) have the LRC implemented (Fig. \ref{LRC}) while the Zeeman qubits have only the standard circuit implemented (Fig. \ref{surface}). The LRC swaps data and ancilla qubits, effectively reinitizating leaked qubits back into the computational subspace. If the magnetic field is stabilized to below $\approx 30$ $\mu$G, the logical error of the Zeeman qubit is better than that of the hyperfine for the scattering rates considered.}
\label{mag}
\end{figure}
\begin{figure}[h]
\includegraphics[width=\columnwidth]{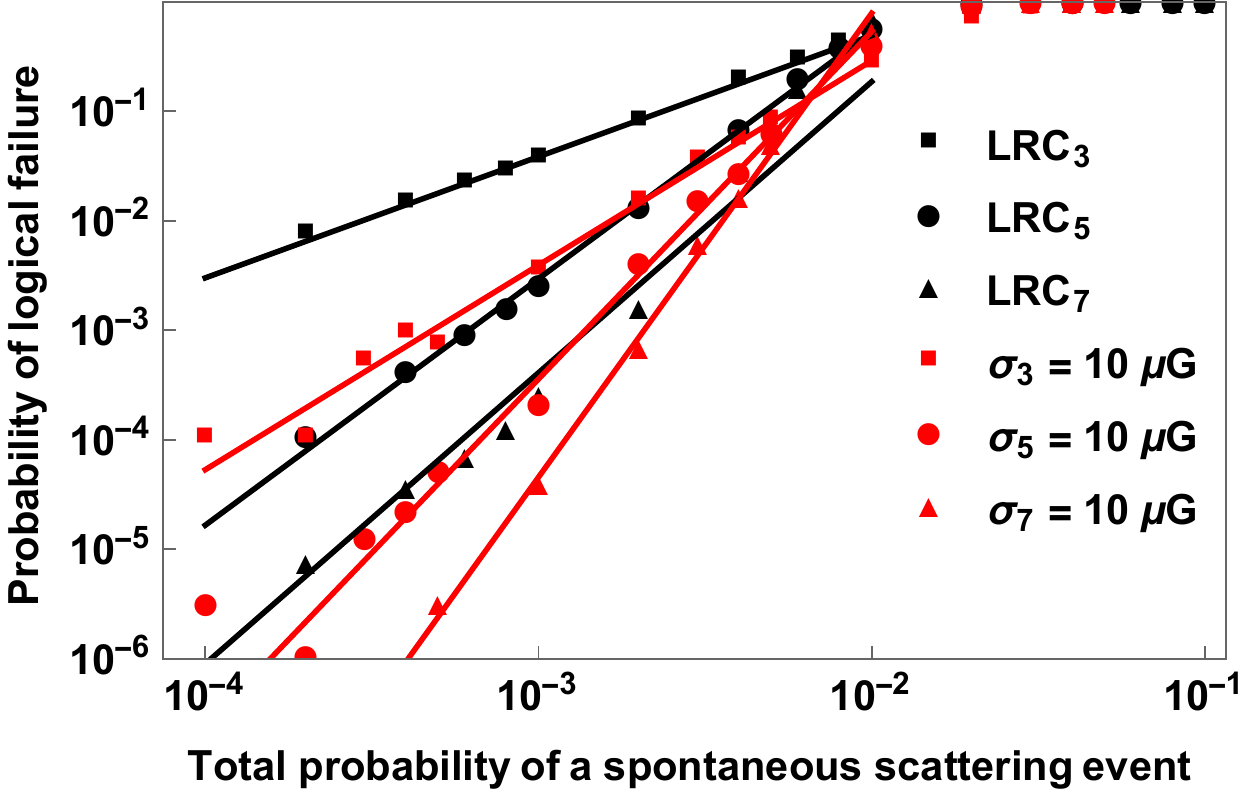}
\caption{Comparison of various distances for hyperfine qubits with the LRC (black) and Zeeman qubits with a magnetic field fluctuations (red) with a standard deviation of 10 $\mu$G, per 2 qubit gate.  The Zeeman qubit yields lower logical error for codes of the same distance.}
\label{dis}
\end{figure}

As expected we found that the success of the Zeeman qubit depended heavily on the stability of the magnetic field. A comparison of the Zeeman and hyperfine qubits at varying magnetic field stabilities is shown in Fig. \ref{mag}. It is clear from this graph that if the magnetic field is not stable enough, the error rate is above threshold and QECC will not help. There is also a stability where the performance of the Zeeman qubit and hyperfine qubit are about equal ($\sigma = 31.62$ $\mu$G), but when the probability of a spontaneous scattering event is low enough, ($\approx10^{-4}$), then the main source of error for the Zeeman qubit is from the magnetic field fluctuation. This base error results in a plateau on the graph were the logical error rate cannot be improved by reducing the scattering. Finally, if the magnetic field can be stabilized to $10$ $\mu$G or less, corresponding to a qubit dephasing error probability per gate of $7.75 \times 10^{-7}$, then the Zeeman qubit produced a lower logical error rate than hyperfine qubit. There did not appear to be a significant improvement of the logical error rate past $10$ $\mu$G for the scattering rates studied. When the field reaches a certain magnitude of stability, the main source of error comes from the spontaneous scattering, which is independent of the magnetic field. Thus the behavior at higher stabilities is more or less the same. 

Using a stability of $10$ $\mu$G, we looked at the behavior of different distance toric codes. Fig. \ref{dis} compares the performance of the two qubits using $d = 3, 5, 7$ codes. It is clear from this that, given the $10$ $\mu$G stability, the Zeeman qubit produces the smaller logical error. With the addition of the LRC, the hyperfine qubits performance was suppressed to that of a lower distance code. The LRC data for $d = 5$ is nearly identical to the standard circuit data for $d = 3$. Similarly, the LRC data for $d = 7$ is comparable to that of the standard circuit data for $d = 5$. A similar behavior was also found by Fowler \cite{fowler}. This behavior suggested a single leakage error may act like two Pauli errors. This is evidence that not all errors are equally damaging. Some errors (such as leakage) can be more harmful to QECC compared to others. Not only do these error require more resources to correct, they suppress the effectiveness of QECC.    

In this sense it is clear that the Zeeman qubit outperforms the hyperfine qubit as it does not require additional circuitry that suppress its performance. However this of course comes with the caveat that the applied magnetic field be stabilized to $\leq 10$ $\mu$G. The existence of a Zeeman qubit in a field of stabilized to $10$ nG has already been physically realized \cite{ruster2016long}.

\section{Conclusions}
Zeeman qubits are prone to more overall physical errors resulting from both magnetic field fluctuations and spontaneous scattering. When the stability of the applied magnetic field is above $30$ $\mu$G, the Zeeman qubit's logical error rate is higher than that of the hyperfine qubit. However, when the magnetic field is stabilized to $\leq 10$ $\mu$G, the logical error rate is suppressed and is less than that of the hyperfine qubit.  

For hyperfine qubits, leakage due to spontaneous scattering is a prominent source of error. These errors are problematic for two reasons: 1) when entangled with other qubits via the CNOT gates, they corrupt the other qubit state and 2) these errors cannot be corrected using standard QEC schemes and require the use of LRCs to correct. For standard QEC schemes, a single physical leakage error has the ability to produce a logical error. This limits the effectiveness of a QECC.  

We have not considered additional physical differences between the hyperfine and Zeeman qubits involving state preparation and measurement.  We have also not considered physical methods of leakage reduction.  For example, perfect polarized $\pi$ light tuned resonant with the S$_{1/2}$, $F=1$ to P$_{1/2}$, $F=1$ transition will remove population from the leaked states for the hyperfine qubit.  The qubit $\ket{1}$ states will have a small probability ($\approx 10^{-4}$) to leak or suffer a bit flip error due to off-resonant $\Delta F = -1$ transitions.  In practice, leakage errors during this procedure will be larger due to imperfect polarization. 

In our study, we also examined the toric code which may be less practical than the planar surface code depending on the layout. Modular architectures could implement the toric code directly \cite{NickersonNatComm2013}, while architectures based on local geometry are better suited to the surface code  \cite{LekitscheSciAdv2017}. For small devices implementing the code in a single ion chain \cite{trout2017simulating}, either the torus or plane would work.  To implement the leakage reduction circuit in the plane, additional circuits on the boundary are necessary to enable the swap.

We have shown that the ideal qubit for near term experiments may not be the ideal qubit for large scale fault-tolerant quantum computation.  Our simulation has centered on trapped ions, but we expect that the design of small quantum systems and error corrected quantum systems will yield different requirements on the qubits.  In particular, for solid-state qubits where the qubits are constructed from multiple components, we expect there will be many interesting tradeoffs between the fidelities of small systems and the overhead required to reach a target logical error.

\section{Acknowledgments}
We thank Kenton Brown and Wes Campbell for useful discussions on Zeeman and hyperfine qubits. The toric surface code simulator was provided by Andrew Cross and Martin Suchara, with permission from IBM. This work was supported by the Office of the Director of National Intelligence - Intelligence Advanced Research Projects Activity through Army Research Office (ARO) contract W911NF-10-1-0231, ARO MURI on Modular Quantum Systems W911NF-16-1-0349, and the National Science Foundation grant PHY-1415461.

\end{document}